\documentclass[12pt] {article}
\usepackage{psfrag}
\usepackage{graphicx}
\usepackage{latexsym,amsfonts}
\usepackage{amssymb}
\begin{document}

\title{Reply on \\ ``Comment on neutrino--mixing interpretation\\ of the GSI 
time anomaly''\\ by C. Giunti\,,\,{\rm nucl--th/0801.4639}}

\author{A. N. Ivanov$^{a}$\thanks{E--mail:
    ivanov@kph.tuwien.ac.at, Tel.: +43--1--58801--14261, Fax:
    +43--1--58801--14299}, R. Reda${^c}$, P. Kienle$^{b,c}$\thanks{E--mail: 
Paul.Kienle@ph.tum.de}}

\date{\today}

\maketitle

\begin{center} {\it $^a$Atominstitut der \"Osterreichischen
    Universit\"aten, Technische Universit\"at Wien, Wiedner
    Hauptstrasse 8-10, A-1040 Wien, \"Osterreich, \\
    $^b$Physik Department, Technische Universit\"at M\"unchen,
    D--85748 Garching, Germany,\\ $^c$Stefan Meyer Institut f\"ur
    subatomare Physik, \"Osterreichische Akademie der Wissenschaften,
    Boltzmanngasse 3, A-1090, Wien, \"Osterreich}
\end{center}

\begin{center}
\begin{abstract}
  Here we reply critically to the comments by Giunti
  (nucl--th/0801.4639) and justify our explanation of the
  experimentally observed periodic interference term in the rate of
  the K--shell electron capture decay of the H--like ions
  ${^{140}}{\rm Pr}^{58+}$ and ${^{142}}{\rm Pm}^{60+}$ as a
  neutrino--flavour mixing.
\end{abstract}
\end{center}

\newpage

\section{Introduction}

According to recent experimental data at GSI \cite{GSI1} on the
K--shell electron capture (EC) decays of the H--like ions
${^{140}}{\rm Pr}^{58+}$ and ${^{142}}{\rm Pm}^{60+}$
\begin{eqnarray}\label{label1}
 &&{^{140}}{\rm Pr}^{58+} \to {^{140}}{\rm Ce}^{58+} + \nu,\nonumber\\
&&{^{142}}{\rm Pm}^{60+} \to {^{142}}{\rm Nd}^{60+} + \nu,
\end{eqnarray}
the rate of the number $N^{EC}_d(t)$ of daughter ions ${^{140}}{\rm
Ce}^{58+}$ or ${^{142}}{\rm Nd}^{60+}$
$$
  \frac{dN^{EC}_d(t)}{dt} = \lambda^{(\rm H)}_{EC}(t)\, N_m(t),\eqno(\rm I.1)
$$ where $N_m(t)$ is the number of mother ions ${^{140}}{\rm
Pr}^{58+}$ or ${^{142}}{\rm Pm}^{60+}$\cite{GSI1} and $\lambda^{(\rm
H)}_{EC}(t)$ is the $EC$--decay rate, is a periodic function, caused
by a periodic time--dependence of the $EC$--decay rate 
$$
  \lambda^{(\rm H)}_{EC}(t) = \lambda^{(\rm H)}_{EC}\,\Big(1 + a_{EC}\,
\cos\frac{2\pi t}{T_d}\Big)\eqno(\rm I.2)
$$
with a period $T_d \simeq 7\,{\rm sec}$ and an amplitude $a_{EC} =
0.20(2)$ \cite{GSI1}.

In our paper \cite{Ivanov} such a periodic time--dependence of the
$EC$--decay rate we have proposed to explain as an interference of
two--neutrino flavours.  We have related the period $T_d$ to the
difference $\Delta m^2_{21} = m^2_2 - m^2_1$ of the squared neutrino
masses $m_2$ and $m_1$
$$ \frac{2\pi}{T_d} = \frac{\Delta m^2_{21}}{4\gamma M_d},\eqno(\rm
 I.3)
$$ where $M_d$ is the mass of the daughter ion and $\gamma = 1.43$ is
a Lorentz factor \cite{GSI1}.

The amplitude $A(I_m \to I_d + \nu)$ of the $EC$--decay $I_m \to I_d +
\nu$, where $I_m$ and $I_d$ are the mother and daughter ions and $\nu$
is a neutrino, which is not detected experimentally \cite{GSI1}, we
define in the form of the coherent sum of the amplitudes $A(I_m \to
I_d + \nu_j)$
$$
 A(I_m \to I_d + \nu) = \sum_{j = 1,2,3}A(I_m \to I_d + \nu_j),\eqno(\rm I.4)
$$ where $\nu_j$ is a neutrino state with mass $m_j$
\cite{Grimus1,PDG06}.

This explanation has been recently criticised by Giunti \cite{Giunti}.
Below we reply on this critique. 

The paper is organised as follows. In section 2 we cite the paper by
Giunti \cite{Giunti} in order to simplify the communication. In
section 3 we give a detailed critical reply on Giunti's critique. We
show that the Eq.(\ref{label9}) and Eq.(\ref{label10}) (see Section 2
and \cite{Giunti}), proposed by Giunti for the definition of the wave
function of a neutrino in the final state of the $EC$--decay and the
amplitude of the $EC$--decay, cannot be used, since they contradict
the main principles of time--dependent perturbation theory and quantum
field theory. In section 4 we give arguments for the description of
the amplitude of the $EC$--decay in the form Eq.(\rm I.4). In Section
5 we apply the procedure, which we used for the analysis of the
$EC$--decay, to the calculation of the time--dependent decay rate of
the $\pi^+ \to \mu^+ + \nu$ decay by defining the amplitude of the
decay as a coherent sum of the amplitudes $\pi^+ \to \mu^+ +
\nu_j$. The obtained result agrees well with the experimental data on
the measurement of the lifetime of the $\pi^+$--meson \cite{PDG06}.
In the Conclusion we summarise our replies.

\section{Comment on neutrino--mixing interpretation of the GSI 
time anomaly by C. Giunti, nucl--th/0801.4639}

The authors of Ref.\cite{Ivanov} calculated the electron capture
process using time--dependent perturbation theory with the effective
time--dependent weak interactions Hamiltonian
\begin{eqnarray}\label{label2}
  \hspace{-0.1in}{\rm H}_W(t ) =  \frac{G_F}{\sqrt{2}}V_{ud}\int d^3x
  [\bar{\psi}_n(x)\gamma^{\mu}(1 -
  g_A\gamma^5) \psi_p(x)]\sum_{j}[U^*_{ej}\bar{\psi}_{\nu_j}(x) 
  \gamma_{\mu}(1 - \gamma^5)\psi_{e^-}(x)]
\end{eqnarray}
with standard notations. They interpreted (see Eq.(3) of Ref.\cite{Ivanov})
\begin{eqnarray}\label{label3}
 A(t) = \sum_kA_k(t)
\end{eqnarray}
as the time--dependent amplitude of the decay
\begin{eqnarray}\label{label4}
I_i \to I_f + \nu_e,
\end{eqnarray}
where
\begin{eqnarray}\label{label5}
A_k(t) = \int^t_0d\tau\,\langle I_f,\nu_k|H_W(\tau)|I_i\rangle
\end{eqnarray}
is the time--dependent amplitude of 
\begin{eqnarray}\label{label6}
I_i \to I_f + \nu_k,
\end{eqnarray}
transitions. Here $I_i$ is the initial ion (${^{140}}{\rm Pr}^{58+}$
or ${^{142}}{\rm Pr}^{60+}$), $I_f$ is the final ion (${^{140}}{\rm
  Ce}^{58+}$ or ${^{142}}{\rm Nd}^{60+}$, respectively), and $\nu_k$
are the massive neutrinos ($k = 1,2,3$).

Regrettably, the amplitude in Eq.(\ref{label3}) does not describe the
decay (4), but a decay in which the final neutrino state is 
\begin{eqnarray}\label{label7}
|\nu\rangle = \sum_k|\nu_k\rangle,
\end{eqnarray}
which is clearly different from an electron neutrino state. Indeed, in
the standard theory of neutrino oscillations (see references of
Ref.\cite{Giunti}) electron neutrinos are described by the state 
\begin{eqnarray}\label{label8}
|\nu_e\rangle = \sum_kU^*_{ek}|\nu_k\rangle,
\end{eqnarray}
where $U$ is the unitary mixing matrix of the neutrino fields in
Eq.(\ref{label2}). More accurately, if the neutrino mass effects in
the interaction processes are taken into account (see references in
\cite{Giunti}), in the time--dependent perturbation theory used in
Ref.\cite{Ivanov} the final electron neutrino in the process (4) is
described by the normalised state
\begin{eqnarray}\label{label9}
|\nu_e(t)\rangle = \frac{\displaystyle \sum_kA_k(t)|\nu_k\rangle}{\displaystyle 
\sqrt{\sum_j|A_j(t)|^2}}.
\end{eqnarray}
The time dependence of this electron neutrino state takes into account
the fact that in time--dependent perturbation theory the final state
of a process is studied during formation.

Using the correct electron neutrino state in Eq.(\ref{label9}), the
decay amplitude is not given by Eq.(\ref{label3}), but by
\begin{eqnarray}\label{label10}
  A(t) =  \frac{\displaystyle \int^t_0d\tau\,\sum_kA^*_k(t)
\langle I_f,\nu_k|H_W(\tau)|I_i\rangle}{\displaystyle 
    \sqrt{\sum_j|A_j(t)|^2}} = \sqrt{\sum_k|A_k(t)|^2}.
\end{eqnarray}
Then, it is clear that the electron capture probability is given by
the incoherent sum over the different channels of massive neutrino
emission. In other words, there is no interference term between
different massive neutrinos contributing to the rates of the electron
capture processes ${^{140}}{\rm Pr}^{58+} \to {^{140}}{\rm Ce}^{58+} +
\nu_e$ and ${^{142}}{\rm Pm}^{60+} \to {^{142}}{\rm Ce}^{60+} +
\nu_e$, as well as all decays and cross sections.

In conclusion, I have shown that neutrino mixing cannot explain the
GSI time anomaly, refuting the claims presented in Ref.\cite{Ivanov}.

\section{Reply on``Comment on neutrino--mixing interpretation of the GSI 
time anomaly'' by C. Giunti, {\rm nucl--th/0801.4639}}

According to Giunti's assertion \cite{Giunti}, the wave function of
 the neutrino in the final state of the $EC$--decay $I_m \to I_d +
 \nu$ should be taken in form Eq.(\ref{label9}) as $|\nu_e(t)\rangle$
 and it should have a non--trivial dependence on time. But
 ``regrettably'' such an assertion contradicts the main principles of
 time--dependent perturbation theory \cite{QM1}--\cite{QM3} and
 quantum field theory \cite{Bjorken}. In order to show this in detail
 we make an excursion to time--dependent perturbation theory
 \cite{QM1}--\cite{QM3}.

\subsubsection*{Analysis of the wave function Eq.(\ref{label9}) in 
 time--dependent perturbation theory}

According to \cite{QM1}--\cite{QM3}, time--dependent perturbation
 theory describes transitions $i \to f$ from the initial stationary
 state $|i\rangle$ with the wave function $\psi^{(0)}_i(0)$ to the
 final stationary state $|f\rangle$ with the wave function
 $\psi^{(0)}_f(0)$, caused by a time--dependent perturbation $H_W(t)$.
 The wave functions $\psi^{(0)}_i(0)$ and $\psi^{(0)}_f(0)$ are
 eigenfunctions of an unperturbed Hamilton operator $H_0$ with
 eigenvalues $E^{(0)}_i$ and $E^{(0)}_f$, respectively. These wave
 functions have no information about a perturbation Hamilton operator
 $H_W(t)$ \cite{QM1}--\cite{QM3}.

The amplitude $a_{mk}(t)$ of the transition of the stationary state
 $|k\rangle$ with the wave function $\psi^{(0)}_k(0)$ to the
 stationary state $|m\rangle$ with the wave function $\psi^{(0)}_m(0)$
 obeys the following differential equation \cite{QM1}--\cite{QM3}
\begin{eqnarray}\label{label11}
   i\,\frac{\partial }{\partial t}a_{mk}(t) = \sum_n
  e^{\textstyle \,i\,(E^{(0)}_m - E^{(0)}_n)t} \langle
  m|H_W(t)|n\rangle\,a_{nk}(t),
\end{eqnarray}
where the matrix element $\langle m|\hat{H}_W(t)|n\rangle$ is defined
by
\begin{eqnarray}\label{label12} 
   \langle m|H_W(t)|n\rangle = \int dv\,
\psi^{(0)\dagger}_m(0) H_W(t) \psi^{(0)}_n(0)
\end{eqnarray}
and $dv$ is an element of a configuration space. Since the interaction
is weak, Eq.(\ref{label11}) can be solved perturbatively.  Keeping the
contributions up to the first order in the Fermi coupling constant
$O(G_F)$, for the coefficients $a_{nk}(t)$ we get the following
expression
\begin{eqnarray}\label{label13} 
  a_{nk}(t) = a^{(0)}_{nk} + a^{(1)}_{nk}(t) = \delta_{nk} + a^{(1)}_{nk}(t)
\end{eqnarray}
where $a^{(0)}_{nk} = \delta_{nk}$ means that a quantum system at
$H_W(t) = 0$ does not change the state. The coefficient
$a^{(1)}_{nk}(t)$ is of order $O(G_F)$. It defines the amplitude of
the $k \to n$ transition, caused by a weak interaction
$H_W(t)$. Substituting Eq.(\ref{label13}) into Eq.(\ref{label11}) we
obtain the amplitude $a^{(1)}_{mk}(t)$ of the $k \to m$ transition,
caused by a weak interaction $H_W(t)$, equal to \cite{QM1}--\cite{QM3}
\begin{eqnarray}\label{label14} 
  a^{(1)}_{mk}(t) = -\,i\int^t_0d\tau\,
 e^{\textstyle \,i\,(E^{(0)}_m - E^{(0)}_k)\tau} \langle m|H_W(\tau)|k\rangle.
\end{eqnarray}
For the transition $i \to f$ we set  $k = i$ and $m = f$ and get
\begin{eqnarray}\label{label15} 
  a^{(1)}_{i\to f}(t) = -\,i\int^t_0d\tau\,
 e^{\textstyle \,i\,(E^{(0)}_f - E^{(0)}_i)\tau} \langle f|H_W(\tau)|i\rangle,
\end{eqnarray}
where $|i\rangle$ and $|f\rangle$ are stationary states with the wave
functions $\psi^{(0)}_i(0)$ and $ \psi^{(0)}_f(0)$, respectively.

Thus, according to standard time--dependent perturbation theory
\cite{QM1}--\cite{QM3}, wave functions of the initial and final states
of the $i \to f$ transition are independent of time. Moreover wave
functions of the initial and final states are eigenfunctions of a
non--perturbed Hamilton $H_0$ and have no information about a
perturbation interaction $H_W(t)$.

Since the time derivative of the wave function $|\nu_e(t)\rangle$ in
Eq.(\ref{label9}) is not equal to zero
\begin{eqnarray}\label{label16} 
 i\,\frac{\partial |\nu_e(t)\rangle}{\partial t} \neq 0,
\end{eqnarray}
the wave function Eq.(\ref{label9}) does not describe a stationary
state and, correspondingly, cannot be used for the calculation of the
amplitude of the $EC$--decay within standard time--dependent
perturbation theory \cite{QM1}--\cite{QM3}. A strong dependence of the
wave function Eq.(\ref{label9}) on a structure of a perturbation
Hamilton operator $H_W(t)$ confirms also the impossibility to use this
wave function for the analysis of the $EC$--decays within standard
time--dependent perturbation theory.

\subsubsection*{Does the wave function Eq.(\ref{label9}) define the  asymptotic neutrino 
state at $t\to\infty\,$?}

Another confirmation of the falseness of the wave function
Eq.(\ref{label9}) as a true wave function for a neutrino in the final
state of the $EC$--decay is the fact that such a wave function does
not describe an ``asymptotic neutrino state'' at $t \to \infty$. In
order to illustrate this assertion we propose to investigate in detail
the application of the wave function Eq.(\ref{label9}) to the
description of the $EC$--decay of the H--like ${^{140}}{\rm Pr}^{58+}$
ion.  For simplicity we can use plane waves for the wave functions of
neutrinos $\nu_j$ with masses $m_j$.  This gives \cite{Ivanov}
\begin{eqnarray}\label{label17}
  \hspace{-0.3in}&&A_j(t) =  \sqrt{3}\,\sqrt{2 M_m 2 E_d}\,{\cal M}_{\rm GT}\,
  \langle \psi^{(Z)}_{1s}\rangle\,(2\pi)^3\delta^{(3)}(\vec{k} + \vec{q}\,)\,U_{ej}
  \sqrt{E_j}\,\frac{\displaystyle
\sin\Big(\frac{\Delta E_j}{2}\,t\Big)
    }{\displaystyle \Big(\frac{\Delta E_j}{2}\Big)}\,
e^{\textstyle\,+i\,\frac{\Delta E_j(k)}{2} t},\nonumber\\
\hspace{-0.3in}&&|A_j(t)|^2 = \Big[\sqrt{3}\,\sqrt{2 M_m 2 E_d}\,{\cal M}_{\rm GT}\,
  \langle \psi^{(Z)}_{1s}\rangle\,(2\pi)^3\delta^{(3)}(\vec{k} + \vec{q}\,)\Big]^2
\,|U_{ej}|^2\,E_j\,\frac{\displaystyle
\sin^2\Big(\frac{\Delta E_j}{2}\,t\Big)
    }{\displaystyle \Big(\frac{\Delta E_j}{2}\Big)^2},
\end{eqnarray}
where $\vec{k}$ and $\vec{q}$ are momenta of the neutrino $\nu_j$ and
the daughter ion, respectively, and
\begin{eqnarray}\label{label18}
  \Delta E_j = \sqrt{k^2 + m^2_j} + E_d(k)  - M_m,
\end{eqnarray}
where $E_d(k)$ and $M_m$ are an energy and mass of the daughter and
mother ions, respectively. The wave function of Eq.(\ref{label9}) with
$A_j(t)$, defined by Eq.(\ref{label17}), takes the form
\begin{eqnarray}\label{label19}
  |\nu_e(t)\rangle = \frac{\displaystyle \sum_j U_{ej}
    \sqrt{E_j}\,\frac{\displaystyle \sin\Big(\frac{\Delta
    E_j}{2}\,t\Big) }{\displaystyle \Big(\frac{\Delta E_j}{2}\Big)}\,
    e^{\textstyle\,+\,i\,\frac{\Delta E_j}{2}
    t}\,|\nu_j\rangle}{\displaystyle \sqrt{\sum_j
    |U_{ej}|^2\,E_j\,\frac{\displaystyle \sin^2\Big(\frac{\Delta
    E_j}{2}\,t\Big) }{\displaystyle \Big(\frac{\Delta
    E_j}{2}\Big)^2}}}.
\end{eqnarray}
Since, according to Giunti \cite{Giunti} (see also Section 2), the
wave function Eq.(\ref{label19}) describes a neutrino state at any
finite time $t$, it should also define the asymptotic neutrino state,
calculated at $t \to \infty$, related to an observable detectable
neutrino state \cite{Bjorken}. Using the relations
\begin{eqnarray}\label{label20}
  \frac{\displaystyle \sin\Big(\frac{\Delta E_j}{2}\,t\Big)
  }{\displaystyle \Big(\frac{\Delta E_j}{2}\Big)} 
\stackrel{t\to \infty}{\longrightarrow} 2\pi\,\delta(\Delta E_j)\quad,\quad
 \frac{\displaystyle
    \sin^2\Big(\frac{\Delta E_j}{2}\,t\Big)
  }{\displaystyle \Big(\frac{\Delta E_j}{2}\Big)^2} 
\stackrel{t\to \infty}{\longrightarrow}  2\pi\,t\,\delta(\Delta E_j)
\end{eqnarray}
and energy conservation we get
\begin{eqnarray}\label{label21}
  |\nu_e(t)\rangle  \stackrel{t\to \infty}{\longrightarrow} 
\sqrt{\frac{2\pi}{t}}\,\frac{\displaystyle \sum_j U_{ej}
    \delta(\Delta E_j)\,|\nu_j\rangle}{\displaystyle 
    \sqrt{\sum_j |U_{ej}|^2\delta(\Delta E_j)}} = O\Big(\frac{1}{\sqrt{t}}\Big).
\end{eqnarray}
Since the r.h.s. of Eq.(\ref{label21}) vanishes in the limit $t \to
\infty$, the wave function Eq.(\ref{label9}) describes no ``asymptotic
neutrino'' state, which can be detected \cite{Bjorken}.

\subsubsection*{Analysis of  the amplitude Eq.(\ref{label10}) and the wave function 
Eq.(\ref{label9}) for the $EC$--decay with the neutrino $\nu_e$,
treated as an elementary particle}

The incorrectness of the relation Eq.(\ref{label10}) becomes obvious
if one treats the neutrino $\nu_e$ in the final state of the $I_m \to
I_d + \nu_e$ decay as an elementary particle.  According to Giunti
\cite{Giunti}, the wave function of the neutrino $\nu_e$ should be
taken in the form
\begin{eqnarray}\label{label22}
  |\nu_e(t)\rangle  = \frac{A(t)\,|\nu_e\rangle}{\sqrt{|A(t)|^2}},
\end{eqnarray}
where $A(t)$ is 
\begin{eqnarray}\label{label23}
A(t) = \int^t_0d\tau\,\langle I_d,\nu_e|H_W(\tau)|I_m\rangle
\end{eqnarray}
In accordance with Eq.(\ref{label10}) (see Section 2 and
\cite{Giunti}), the amplitude $A(t)$ of the $I_m \to I_d + \nu_e$
should be defined by
\begin{eqnarray}\label{label24}
A(t) = \sqrt{|A(t)|^2} = |A(t)|.
\end{eqnarray}
This means that the amplitude $A(t)$ is positive and has no imaginary
part. Unlike the result, obtained following Giunti's prescription
Eq.(\ref{label10}), the direct calculation of the amplitude $A(t)$ of
the $I_m \to I_d + \nu_e$ decay gives the expression \cite{Ivanov}
\begin{eqnarray}\label{label25}
  A(t) =i\,\sqrt{3}\,\sqrt{2 M_m 2E_d E_{\nu}}\,{\cal M}_{\rm GT}\,
\langle \psi^{(Z)}_{1s}\rangle\,(2\pi)^3\,
  \delta^{(3)}(\vec{q} + \vec{k}_{\nu})\,\frac{\displaystyle
    \sin\Big(\frac{\Delta E_{\nu}}{2}\,t\Big)
  }{\displaystyle \Big(\frac{\Delta E_{\nu}}{2}\Big)}\,
  e^{\textstyle\,+\,i\,\frac{\Delta E_{\nu}}{2} t},
\end{eqnarray}
where $\Delta E_{\nu} = E_{\nu}(k_{\nu}) + E_d(k_{\nu}) - M_m$. It is
seen that unlike Eq.(\ref{label24}), imposed by Giunti's prescription
Eq.(\ref{label10}), the amplitude Eq.(\ref{label25}) has a nontrivial
imaginary phase.

Substituting Eq.(\ref{label25}) into Eq.(\ref{label22}) we get the
wave function
\begin{eqnarray}\label{label26}
  |\nu_e(t)\rangle = i\,{\rm sign}\!\!\left[\frac{\displaystyle
\sin\Big(\frac{\Delta E_{\nu}}{2}\,t\Big) }{\displaystyle
\Big(\frac{\Delta E_{\nu}}{2}\Big)}\right]
e^{\textstyle\,+\,i\,\frac{\Delta E_{\nu}}{2} t}\,|\nu_e\rangle,
\end{eqnarray}
where ${\rm sign}[f(t)] = \pm 1$ for $f(t)\gtrless 0$.

Thus, the wave function Eq.(\ref{label26}), constructed in accordance
with Giunti's prescription Eq.(\ref{label9}), does not describe a
stationary neutrino state, therefore it cannot be used for the
calculation of the amplitude of the $I_m \to I_d + \nu_e$ decay in
standard time--dependent perturbation theory \cite{QM1}--\cite{QM3}
with the neutrino $\nu_e$, treated as an elementary particle.

\section{Amplitudes of decays with undetected neutrinos}

The problem, which we discuss in this section, concerns the definition
of the amplitudes of two--body weak decays $I_m \to I_d + \nu$ with
undetected neutrinos, where $I_m$ and $I_d$ are the initial and final
nuclear states. The experimental analysis of the reaction $I_m \to I_d
+ \nu$ contains the preparation of the initial nuclear state $I_m$,
which should be the H--like ion, and the detection of the final
nuclear state $I_d$, which is a bare nucleus.  The neutrino is not
detected.

According to a modern theory of neutrino physics \cite{Grimus1},
neutrinos can be detected only in the states with definite leptonic
flavours $|\nu_e\rangle$, $|\nu_{\mu}\rangle$ or $|\nu_{\tau}\rangle$,
which are superpositions of the neutrino states $|\nu_j\rangle$ with
masses $m_j$
\begin{eqnarray}\label{label27}
|\nu_{\alpha}\rangle = \sum_{j = 1,2,3} U^*_{\alpha j}|\nu_j\rangle,
\end{eqnarray}
where $\alpha = e, \mu$ and $\tau$ for the electron $e^-$, the muon
$\mu^-$ and $\tau$--lepton $\tau^-$, respectively, $U_{\alpha j}$ are
elements of the $3\times 3$ unitary matrix $U$
\cite{Grimus1,PDG06}. The neutrinos $\nu_j$ are not detectable
directly, since they have no definite leptonic flavour \cite{Grimus1}.

The weak interactions of neutrinos $\nu_j$ with leptons and
hadrons (or nuclei) are defined by the current $\times$ current
interactions
\begin{eqnarray}\label{label28}
H^{(h)}_W(t) &=& \frac{G_F}{\sqrt{2}}\,V^{(h)}_{\rm CKM}\sum_{\alpha}\int
d^3x\,J^{(\rm h)}_{\rho}(x)J^{\rho}_{(\alpha)}(x) = \nonumber\\
&=&\frac{G_F}{\sqrt{2}}\,V^{(h)}_{\rm CKM}\sum_{\alpha,j}U_{\alpha j}\int
d^3x\,J^{(\rm h)}_{\rho}(x)\,[\bar{\psi}_{\nu_j}(x)\gamma^{\rho}(1 -
\gamma^5) \psi_{\alpha}(x)],
\end{eqnarray}
caused by the W--boson exchange, where $G_F$ is the Fermi weak
constant, $J^{(\rm h)}_{\rho}(x)$ is a charged hadronic current,
$V^{(h)}_{\rm CKM}$ is a matrix element of the
Cabibbo--Kobayashi--Maskawa (CKM) matrix dependent on the structure of
the hadronic current \cite{PDG06} and $J^{\rho}(x)$ is the charged
leptonic current, defined by \cite{Grimus1}
\begin{eqnarray}\label{label29}
J^{\rho}_{(\alpha)}(x) = \sum_{j = 1,2,3}U_{\alpha
j}\bar{\psi}_{\nu_j}(x)\gamma^{\rho}(1 - \gamma^5)\psi_{\alpha}(x),
\end{eqnarray}
where $\psi_{\nu_j}(x)$ and $\psi_{\alpha}(x)$ are operators of the
neutrino fields $\nu_j$ with masses $m_j$ and lepton fields $-$ the
electron $e^-$, the muon $\mu^-$ and the $\tau$--lepton $\tau^-$ for
$\alpha = e,\mu$ and $\tau$, respectively.

Using the Hamilton operator of the weak interactions
Eq.(\ref{label28}) we can solve the following problem
\cite{Grimus2,Grimus3}. Let a neutrino $\nu$, emitted in the reaction
$I_m \to I_d + \nu$, be used for the subsequent reaction $\nu + X\to Y
+ e^-$, where $X$ and $Y$ are two hadronic or nuclear states. In this
case, according to \cite{Grimus2,Grimus3}, the amplitude of the
transition $I_m \to I_d + \nu \Longrightarrow \nu + X\to Y + e^-$ can
be defined to the second order of time--dependent perturbation theory
as
\begin{eqnarray}\label{label30}
\hspace{-0.3in}&&A_{I_m \to Y + e^-}(t) = -i\sum_{j = 1,2,3}\int^t_0dt''\langle e^-Y|H_W|
X\nu_j\rangle\,e^{i(E_e + E_Y - E_X - E_j)t''} A(I_m \to I_d + \nu_j)(t) =\nonumber\\
\hspace{-0.3in}&&= -\sum_{j = 1,2,3}\int^t_0dt''\langle e^-Y|H_W|
X\nu_j\rangle\,e^{i(E_e + E_Y - E_X - E_j)t''}\int^{t''}_0dt'\,
\langle \nu_j I_d|H_W|I_m\rangle\,
e^{\,i(E_j + E_d - E_m)t'}.\nonumber\\
\hspace{-0.3in}&&
\end{eqnarray}
where $E_e$, $E_Y$, $E_X$ and $E_j$ are the energies of the electron,
the hadronic (or nuclear) states $Y$ and $X$ and the neutrino,
respectively. The amplitudes of the $I_m \to I_d + \nu_j$ transitions
are defined to the first order of time--dependent perturbation theory
as \cite{Grimus2,Grimus3}
\begin{eqnarray}\label{label31}
A(I_m \to I_d +  \nu_j)(t)= -i\int^t_0dt'\,\langle \nu_j,I_d|H_W|I_m\rangle\,
e^{\,i(E_j + E_d - E_m)t'},
\end{eqnarray}
where $E_m$ and $E_d$ are energies of the initial and final hadronic
(nuclear) states and $E_j$ is a neutrino energy.

Thus, the amplitude of the transition $I_m \to I_d + \nu
\Longrightarrow \nu + X\to Y + e^-$ is a coherent sum of the
amplitudes of the transitions $I_m \to I_d + \nu_j \Longrightarrow
\nu_j + X\to Y + e^-$. As a result the rate of the transition $I_m \to
I_d + \nu \Longrightarrow \nu + X\to Y + e^-$, defined by $|A_{I_m \to
Y + e^-}(t)|^2$, should contain both the squared absolute values of
the amplitudes of the $I_m \to I_d + \nu_j \Longrightarrow \nu_j +
X\to Y + e^-$ transitions and the interference terms
\cite{Grimus2,Grimus3}.

Now let us consider the $I_m \to I_d + \nu_e$ decay. Using the
definition of the wave function of the electronic neutrino
Eq.(\ref{label27}), the weak interaction Hamilton operator
Eq.(\ref{label28}) and standard time--dependent perturbation theory
\cite{QM1}--\cite{QM3}, for the amplitude of the $I_m \to I_d + \nu_e$
decay, dependent on time $t$, we obtain the following expression
\begin{eqnarray}\label{label32}
A(I_m \to I_d + \nu_e)(t) &=& \sum_j \langle \nu_e|\nu_j\rangle\,A(I_m
\to I_d + \nu_j)(t) = \sum_j U_{ej}A(I_m \to I_d + \nu_j)(t) =
\nonumber\\ &=& -i\sum_j\int^t_0dt'\,U_{ej}\langle
\nu_j,I_d|H_W|I_m\rangle\, e^{\,i(E_j + E_d - E_m)t'}.
\end{eqnarray}
Thus, the amplitude of the $I_m \to I_d + \nu_e$ decay is a coherent
sum of the amplitudes $I_m \to I_d + \nu_j$ with a weight $|U_{e
j}|^2$.  The decay rate $\lambda(t)$ is defined by
\begin{eqnarray}\label{label33}
\lambda(t) \propto \int |A(I_m \to I_d +  \nu_e)(t)|^2\,d\rho,
\end{eqnarray}
where $d\rho$ is an element of a phase volume of the final state. The
decay rate contains both the contributions of the squared absolute
values of the amplitudes of the transitions $I_m \to I_d + \nu_j$ and
the interference terms.

For the $EC$--decay of the H--like ${^{140}}{\rm Pr}^{58+}$ ion
${^{140}}{\rm Pr}^{58+} \to {^{140}}{\rm Ce}^{58+} + \nu_e$,
calculated for the mixing angle $\theta_{13} = 0$ \cite{PDG06}, the
decay rate is equal to
\begin{eqnarray}\label{label34}
  \lambda^{(\rm H)}_{EC}(t) = \Big(1 - \frac{1}{2}\,\sin^2(2\theta_{12})\Big)\,
\lambda^{(\rm H)}_{EC}\,\Big\{1 + a_{EC}
  \cos\Big(\frac{\Delta m^2_{21}}{4 M_d}t\Big)\Big\}.
\end{eqnarray}
where $\lambda^{(\rm H)}_{EC}$ has been calculated in \cite{AIvanov}
(see also \cite{Ivanov}) and reads
\begin{eqnarray}\label{label35}
 \lambda^{\rm (H)}_{EC}= \frac{1}{2 F + 1}\,\frac{3}{2}
  |{\cal M}_{\rm GT}|^2 |\langle \psi^{(Z)}_{1s}\rangle|^2
  \frac{Q^2_{\rm H}}{\pi}.
\end{eqnarray}
The amplitude $a_{EC}$ of a periodic dependence of the decay rate is
\begin{eqnarray}\label{label36}
a_{EC} = \frac{\sin^2(2\theta_{12})}{1 +
\cos^2(2\theta_{12})}\,e^{\,-\delta^2\Delta^2\vec{k}_{21}}.
\end{eqnarray}
For the averaged over time decay rate $\langle \lambda^{(\rm
H)}_{EC}(t)\rangle$ we obtain the following expression
\begin{eqnarray}\label{label37}
  \langle \lambda^{(\rm H)}_{EC}(t)\rangle = \Big(1 - \frac{1}{2}\,
\sin^2(2\theta_{12})\Big)\,
\lambda^{(\rm H)}_{EC}.
\end{eqnarray}
According to \cite{AIvanov}, the appearance of the factor $1 -
\frac{1}{2}\,\sin^2(2\theta_{12}) = 0.57$, calculated for the
experimental value of the mixing angle $\theta_{12} = 33.9$ degrees
\cite{PDG06}, contradicts the experimental data by GSI on the ratios
of the $EC$ and $\beta^+$ decays of the H--like ${^{140}}{ \rm
Pr}^{58+}$ and He--like ${^{140}}{ \rm Pr}^{57+}$ ions \cite{GSI2}.

This shows that unlike Giunti's assertion (see a discussion above
Eq.(\ref{label8}) in Section 2 and Ref.\cite{Giunti}) one cannot use
the wave function $|\nu_e\rangle = \sum_j U^*_{e j}|\nu_j\rangle$ for
the analysis of the $EC$--decay of the H--like ${^{140}}{\rm
Pr}^{58+}$ ion.

Thus, the amplitude of the two--body weak decay $I_m \to I_d + \nu$
 with neutrinos in the final state should be taken in the form of a
 coherent sum of the amplitudes $I_m \to I_d + \nu_j$. For the
 undetected neutrino and unfixed leptonic flavour of the neutrino
 state the amplitude of the $I_m \to I_d + \nu$ is equal to
\begin{eqnarray}\label{label38}
 A(I_m \to I_d + \nu) = \sum_jA(I_m \to I_d + \nu_j).
\end{eqnarray}
As has been shown in \cite{Ivanov}, the decay rate of the $EC$--decay
of the H--like ${^{140}}{\rm Pr}^{58+}$ ion ${^{140}}{\rm Pr}^{58+}
\to {^{140}}{\rm Ce}^{58+} + \nu_e$, calculated for the mixing angle
$\theta_{13} = 0$ \cite{PDG06}, is
\begin{eqnarray}\label{label39}
  \lambda^{(\rm H)}_{EC}(t) = \lambda^{(\rm H)}_{EC}\,\Big\{1 + a_{EC}
\cos\Big(\frac{\Delta m^2_{21}}{4 M_d}t\Big)\Big\},
\end{eqnarray}
where $\lambda^{(\rm H)}_{EC}$ is given by Eq.(\ref{label35}). For the
averaged over time decay rate $\langle \lambda^{(\rm
H)}_{EC}(t)\rangle$ we obtain the following expression \cite{AIvanov}
\begin{eqnarray}\label{label40}
  \langle \lambda^{(\rm H)}_{EC}(t)\rangle = \lambda^{(\rm H)}_{EC}.
\end{eqnarray}
which describes well the experimental data by GSI on the ratios of the
$EC$ and $\beta^+$ decays of the H--like ${^{140}}{ \rm Pr}^{58+}$ and
He--like ${^{140}}{ \rm Pr}^{57+}$ ions \cite{GSI2}.

\section{On time--dependence of  $\pi^+ \to \mu^+ + \nu$ decay rate}

In this section we calculate the decay rate $\lambda_{\pi^+}(t)$ of
the $\pi^+$--meson decay $\pi^+ \to \mu^+ + \nu$. Following standard
theory of weak interactions, standard time--dependent perturbation
theory and using a coherent contribution of the decay channels $\pi^+
\to \mu^+ + \nu_j$ \cite{Ivanov}, for the amplitude of the $\pi^+ \to
\mu^+ + \nu$ decay we obtain the following expression
\begin{eqnarray}\label{label41}
  \hspace{-0.3in}&&A(\pi^+ \to\mu^+\, \nu) = \sum_j  A(\pi^+ \to\mu^+ + \nu_j) =  
  G_FV_{ud}F_{\pi}m_{\mu}\sum_jU_{\mu j}
  i\int^t_0 d\tau\,
  (2\pi \delta^2)^{3/2}\nonumber\\
  \hspace{-0.3in} &&\times\int \frac{d^3k}{(2\pi)^3}\,
  e^{\,-\frac{1}{2}\,\delta^2\,(\vec{k} - \vec{k}_j)^2}
  u^{\dagger}_{\nu_j}(\vec{k},\sigma)
  (1 + \gamma^5) v_{\mu}(\vec{p}_+,\sigma_+)\,e^{\,-\,i(\vec{k} + \vec{p}_+)\cdot \vec{r} 
    + i(E_j(\vec{k}\,) + E_+(\vec{p}_+) - m_{\pi})\tau} = \nonumber\\
  \hspace{-0.3in}&&= - 2 m_{\mu} \sqrt{2 m_{\pi}} G_FV_{ud}F_{\pi}\sum_jU_{\mu j} 
\sqrt{E_j}
  \,
  e^{\,-\frac{1}{2}\,\delta^2\,(\vec{p}_+ + \vec{k}_j)^2} \,\frac{\displaystyle 
    \sin\Big(\frac{\Delta E_j}{2}t\Big)}{\displaystyle 
    \Big(\frac{\Delta E_j}{2}\Big)}\,e^{\,i\Delta E_j\frac{t}{2}},
\end{eqnarray}
where $F_{\pi} = 92.4\,{\rm MeV}$ is the $\pi^+$--meson leptonic
constant \cite{PDG06}, $m_{\pi} = 139.57\,{\rm MeV}$ and $m_{\mu} =
105.66\,{\rm MeV}$ are the pion and muon masses, respectively, $\Delta
E_j = E_j(\vec{p}_+) + E_+(\vec{p}_+) - m_{\pi}$, $k$ and $p_+$ are
4--momenta of the neutrino and the $\mu^+$--meson, respectively. The
squared absolute value of the amplitude is equal to
\begin{eqnarray}\label{label42}
  \hspace{-0.3in}&&|A(\pi^+ \to\mu^+ \,\nu)|^2 = 8 m^2_{\mu} m_{\pi} G^2_F
  |V_{ud}|^2 F^2_{\pi}\,(2\pi \delta^2)^3\nonumber\\
  \hspace{-0.3in}&&\times\,\Bigg\{\sum_j|U_{\mu j}|^2 E_j
  e^{\,-\,\delta^2\,(\vec{p}_+ + \vec{k}_j)^2}\,\frac{\displaystyle 
    \sin^2\Big(\frac{\Delta E_j}{2}t\Big)}{\displaystyle 
    \Big(\frac{\Delta E_j}{2}\Big)^2} + \sum_{i < j}
  2U^*_{\mu i}U_{\mu j}\sqrt{E_iE_j}\nonumber\\
  \hspace{-0.3in}&&\times\,
  e^{\,-\,\delta^2\,\Big(\vec{p}_+ + \frac{\vec{k}_i + \vec{k}_j}{2}\Big)^2}\,
  e^{\,-\,\delta^2\,\Big( \frac{\vec{k}_i - \vec{k}_j}{2}\Big)^2}\,
  \frac{\displaystyle 
    \sin\Big(\frac{\Delta E_i}{2}t\Big)}{\displaystyle 
    \Big(\frac{\Delta E_i}{2}\Big)}\,
  \,\frac{\displaystyle 
    \sin\Big(\frac{\Delta E_j}{2}t\Big)}{\displaystyle 
    \Big(\frac{\Delta E_j}{2}\Big)}\nonumber\\
  \hspace{-0.3in}&&\times\,
  \cos\Big(\frac{E_i - E_j}{2}\,t\Big)\Bigg\}.
\end{eqnarray}
Following \cite{Ivanov}, we obtain the neutrino spectrum. For this aim
we integrate over the phase volume of the positron. This gives
\cite{Ivanov}
\begin{eqnarray}\label{label43}
  &&N_{\nu}(t) = \frac{1}{2 m_{\pi}}\int |A(\pi^+\to\mu^+\,\nu)|^2\,
  \frac{d^3p_+}{(2\pi)^3 2 E_+(\vec{p}_+)} = 2 m^2_{\mu} G^2_F
  |V_{ud}|^2 F^2_{\pi}\,(\pi \delta^2)^{3/2}\nonumber\\
  \hspace{-0.3in}&&\times\,\Bigg\{\sum_j\frac{|U_{\mu
  j}|^2}{E_+(\vec{k}_j)}E_j(\vec{k}_j)\, \frac{\displaystyle
  \sin^2\Big(\frac{\Delta E_j(\vec{k}_j)}{2}t\Big)}{\displaystyle
  \Big(\frac{\Delta E_j(\vec{k}_j)}{2}\Big)^2} + \sum_{i < j}
  \frac{2U^*_{\mu i}U_{\mu j}}{\textstyle E_+\Big(\frac{\vec{k}_i +
  \vec{k}_j}{2}\Big)} \sqrt{E_i(\vec{k}_i)E_j(\vec{k}_j)}\nonumber\\
  \hspace{-0.3in}&&\times\,
  e^{\,-\,\delta^2\,\Delta^2 \vec{k}_{ij}}\,
  \frac{\displaystyle 
    \sin\Big(\frac{\Delta E_i(\vec{k}_i)}{2}t\Big)}{\displaystyle 
    \Big(\frac{\Delta E_i(\vec{i}_i)}{2}\Big)}\,
  \,\frac{\displaystyle 
    \sin\Big(\frac{\Delta E_j(\vec{k}_j)}{2}t\Big)}{\displaystyle 
    \Big(\frac{\Delta E_j(\vec{k}_j)}{2}\Big)}\,
  \cos\Big(\frac{E_i(\vec{k}_i) - E_j(\vec{k}_j)}{2}\,t\Big)\Bigg\}.
\end{eqnarray}
The $\pi^+$--meson decay rate $\lambda_{\pi^+}(t)$ is equal to
\cite{Ivanov}
\begin{eqnarray}\label{label44}
  \hspace{-0.3in}\lambda_{\pi^+}(t) = \int \frac{d^3k}{(2\pi)^3 2 E_{\nu}}
  \frac{N_{\nu}(t)}{t(\pi \delta^2)^{3/2}} = \lambda_{\pi^+}\Big(1 + \sum_{i < j}
  2U^*_{\mu i}U_{\mu j}\,e^{\,- \delta^2 \Delta^2 \vec{k}_{ij}}\,
\cos\Big(\frac{2\pi t}{T_{ij}}\Big)\Big),
\end{eqnarray}
where we have denoted
\begin{eqnarray}\label{label45}
 T_{ij} = \frac{4\pi }{m_{\pi}}\,\frac{m^2_{\pi} - m_{\mu}}{m^2_i - m^2_j}
\end{eqnarray}
and
\begin{eqnarray}\label{label46}
  \lambda_{\pi^+} = \frac{G^2_F |V_{ud}|^2}{4\pi}\,F^2_{\pi} m^2_{\mu} m_{\pi}
\Big(1 - \frac{m^2_{\mu}}{m^2_{\pi}}\Big)^2 = 2.49\times 10^{-14}\,{\rm MeV}.
\end{eqnarray}
The theoretical value $\lambda_{\pi^+} = 2.49\times 10^{-14}\,{\rm
MeV}$ agrees well with the experimental one $\lambda^{\exp}_{\pi^+} =
2.53 \times 10^{-14}\,{\rm MeV}$ \cite{PDG06}.

The time--dependence of the decay rate $\lambda_{\pi^+}(t)$ is defined
by
\begin{eqnarray}\label{label47}
  \lambda_{\pi^+}(t) = \lambda_{\pi^+}\Big(1 &-& \sin
2\theta_{12}\,\cos^2\theta_{23}\,e^{\,-\,\delta^2\, \Delta^2
\vec{k}_{21}}\,\cos\Big(\frac{2\pi t}{T_{21}}\Big)\nonumber\\ &-& \sin
\theta_{12}\,\sin2\theta_{23}\,e^{\,-\,\delta^2\, \Delta^2
\vec{k}_{31}}\,\cos\Big(\frac{2\pi t}{T_{31}}\Big)\nonumber\\ &+& \cos
\theta_{12}\,\sin2\theta_{23}\,e^{\,-\,\delta^2\, \Delta^2
\vec{k}_{32}}\,\cos\Big(\frac{2\pi t}{T_{32}}\Big)\Big),
\end{eqnarray}
where we have used matrix elements $U_{\mu j}$ \cite{PDG06} and set
$\theta_{13} = 0$ \cite{PDG06,Fogli}. The periods $T_{ij}$ are equal
to
\begin{eqnarray}\label{label48}
T_{21} &=& \frac{4\pi}{m_{\pi}}\,\frac{m^2_{\pi} -
  m^2_{\mu}}{\Delta m^2_{21}} = 1.11\times 10^{-3}\,{\rm
  sec},\nonumber\\ T_{31} &=&
\frac{4\pi}{m_{\pi}}\,\frac{m^2_{\pi} - m^2_{\mu}}{\Delta m^2_{31}} =
2.04\times 10^{-4}\,{\rm sec},\nonumber\\ T_{32} &=&
\frac{4\pi}{m_{\pi}}\,\frac{m^2_{\pi} - m^2_{\mu}}{\Delta m^2_{32}} =
2.04\times 10^{-4}\,{\rm sec},
\end{eqnarray}
were we have set $\Delta m^2_{31} \simeq \Delta m^2_{32} = 2.40\times
10^{-3}\, {\rm eV}^2 $ \cite{PDG06,Fogli}, $m_{\pi} = 139.57\,{\rm
MeV}$ and $m_{\mu} = 105.66\,{\rm MeV}$ \cite{PDG06}.  The periods of
oscillations are much greater than the lifetime $\tau_{\pi^+} =
2.60\times 10^{-8}\,{\rm sec}$ \cite{PDG06}.

\section{Conclusion}

Concluding our replies on Giunti's critique we argue that

\begin{itemize}
\item {\bf The amplitude of the $EC$--decay $I_m \to I_d + \nu$ with
an undetected neutrino can be written in the form of the coherent sum
of the amplitudes $I_m \to I_d + \nu_j$ of the $EC$--decays: \[A(I_m
\to I_d + \nu) = \sum_j A(I_m \to I_d + \nu_j).\]}
\item {\bf The wave function of the neutrino in the final state of the
    $EC$--decay $I_m \to I_d + \nu$ cannot be taken in the form of
    Eq.(\ref{label8}), as it is proposed by Giunti \cite{Giunti}. This
    leads to the contradiction with the experimental data on the ratios
    of the $EC$ and $\beta^+$ decays of the H--like and He--like ions
    at GSI \cite{GSI2}.}
\item {\bf Since the wave function Eq.(\ref{label9}), proposed by
Giunti as a wave function of a neutrino in the final state of the
$EC$--decay of the H--like ion \cite{Giunti}, does not describe a
stationary neutrino state, required by time--dependent perturbation
theory, it cannot be used as a wave function of a neutrino in the
final state of the $EC$--decay $I_m \to I_d + \nu$ of the H--like
ion.}
\item {\bf The falseness of the wave function Eq.(\ref{label9}),
proposed by Giunti as a wave function of a neutrino in the final state
of the $EC$--decay of the H--like ion \cite{Giunti}, is confirmed also
by the failure of this wave function to describe an asymptotic
neutrino state, related to an observable detectable neutrino state
\cite{Bjorken}.}
\item {\bf The relation Eq.(\ref{label10}), proposed by Giunti for the
definition of the amplitude of the $EC$--decay of the H--like ions, is
incorrect, since it is based on the use of the incorrect wave function
Eq.(\ref{label9}).}
\item {\bf The direct calculation shows also that Eq.(\ref{label10})
  cannot be used for the definition of the amplitude of the $I_m \to
  I_d + \nu_e$ decay even if the neutrino $\nu_e$ is an elementary
  particle.}
\item {\bf The calculation of the $\pi^+ \to \mu^+ + \nu$ decay rate
shows that the definition of the amplitude of the $\pi^+ \to \mu^+ +
\nu$ decay in the form of the coherent sum of the amplitudes of the
decay channels $\pi^+ \to \mu^+ + \nu_j$ \[A(\pi^+ \to \mu^+ + \nu) =
\sum_j A(\pi^+ \to \mu^+ + \nu_j)\] leads to the correct description
of the decay rate $\lambda_{\pi^+}(t)$, agreeing well with the
experimental data on the lifetime of the $\pi^+$--meson \cite{PDG06},
and with a periodic time--dependence \cite{Ivanov}.  Since the periods
of variation of the time--dependent terms are much greater than the
lifetime of the $\pi^+$--meson, such a time--dependence cannot be
measured.}
\end{itemize}

\end{document}